\documentclass[longauth]{aa}  
\usepackage{tabularx}
\usepackage{graphicx}
\usepackage{txfonts}
\usepackage{ragged2e}
\usepackage{booktabs}
\usepackage[colorlinks = true,
            linkcolor = blue,
            urlcolor  = blue,
            citecolor = blue,
            anchorcolor = blue]{hyperref}
\usepackage{url}
\usepackage[encapsulated]{CJK}

\DeclareUnicodeCharacter{05BF}{}
\begin{document} 

   \title{VENUS: When Red meets Blue -- A multiply imaged \emph{Little Red Dot} with an apparent blue companion behind the galaxy cluster Abell 383}

   \author{Miriam Golubchik\inst{1}\thanks{\email{golubmir@post.bgu.ac.il}},
           Lukas J. Furtak\inst{2,3},
           Joseph F. V. Allingham\inst{1},
           Adi Zitrin\inst{1},
           Hollis B. Akins\inst{2,3},
           Vasily Kokorev\inst{2,3},
           Seiji Fujimoto\inst{4,5},
           Abdurro'uf\inst{6},
           Ricardo O. Amor\'{i}n\inst{7},
           Franz E. Bauer\inst{8},
           Rachel Bezanson\inst{9},
           Marusa Brada\v{c}\inst{10,11},
           Larry D. Bradley\inst{12},
           Gabriel B. Brammer\inst{13,14},
           John Chisholm\inst{2,3},
           Dan Coe\inst{12,15,16},
           Christopher J. Conselice\inst{17},
           Pratika Dayal\inst{4,18,19},
           Miroslava Dessauges-Zavadsky\inst{20},
           Jose M. Diego\inst{21},
           Andreas L. Faisst\inst{22},
           Qinyue Fei\inst{4},
           Henry C. Ferguson\inst{12},
           Steven L. Finkelstein\inst{2,3},
           Brenda L. Frye\inst{23},
           Mauro González-Otero\inst{7},
           Jenny E. Greene\inst{24},
           Yuichi Harikane\inst{25},
           Tiger Yu-Yang Hsiao\inst{2,3},
           Kohei Inayoshi\inst{26},
           Yolanda Jim\'enez-Teja\inst{7,27},
           Kirsten Knudsen\inst{28},
           Anton M. Koekemoer\inst{12},
           Ivo Labb\'{e}\inst{29},
           Ray A. Lucas\inst{12},
           Georgios E. Magdis\inst{13,14},
           Jorryt Matthee\inst{30},
           Matteo Messa\inst{31},
           Rohan P. Naidu\inst{32},
           Minami Nakane\inst{25,33},
           Ga\"el Noirot\inst{12},
           Richard Pan\inst{34},
           Casey Papovich\inst{35,36},
           Johan Richard\inst{37},
           Massimo Ricotti\inst{38},
           Luke Robbins\inst{33},
           Daniel P. Stark\inst{39},
           Fengwu Sun\inst{40},
           Tommaso Treu\inst{41},
           Roberta Tripodi\inst{42},
           Eros Vanzella\inst{31},
           Chris Willott\inst{43},
           \and
           Rogier A. Windhorst\inst{44}
          }

   \institute{Department of Physics, Ben-Gurion University of the Negev, P.O. Box 653, Be'er-Sheva 84105, Israel
   \and
   Cosmic Frontier Center, The University of Texas at Austin, Austin, TX 78712, USA
   \and
   Department of Astronomy, The University of Texas at Austin, 2515 Speedway, Stop C1400, Austin, TX 78712, USA
   \and
   David A. Dunlap Department of Astronomy and Astrophysics, University of Toronto, 50 St. George Street, Toronto, Ontario, M5S 3H4, Canada
   \and
   Dunlap Institute for Astronomy and Astrophysics, 50 St. George Street, Toronto, Ontario, M5S 3H4, Canada
   \and
   Department of Astronomy, Indiana University, 727 East Third Street, Bloomington, IN 47405, USA
   \and
   Instituto de Astrof\'{i}sica de Andaluc\'{i}a (CSIC), Apartado 3004, 18080 Granada, Spain
   \and
   Instituto de Alta Investigaci{\'{o}}n, Universidad de Tarapac{\'{a}}, Casilla 7D, Arica, Chile
   \and
   Department of Physics and Astronomy and PITT PACC, University of Pittsburgh, Pittsburgh, PA 15260, USA
   \and
   University of Ljubljana, Faculty of Mathematics and Physics, Jadranska ulica 19, SI-1000 Ljubljana, Slovenia
   \and
   Department of Physics and Astronomy, University of California Davis, 1 Shields Avenue, Davis, CA 95616, USA
   \and
   Space Telescope Science Institute (STScI), 3700 San Martin Drive, Baltimore, MD 21218, USA
   \and
   Cosmic Dawn Center (DAWN), Copenhagen, Denmark
   \and
   Niels Bohr Institute, University of Copenhagen, Jagtvej 128, Copenhagen, Denmark
   \and
   Association of Universities for Research in Astronomy (AURA), Inc. for the European Space Agency (ESA)
   \and
   Center for Astrophysical Sciences, Department of Physics and Astronomy, The Johns Hopkins University, 3400 N Charles St. Baltimore, MD 21218, USA
   \and
   Jodrell Bank Centre for Astrophysics, University of Manchester, Oxford Road, Manchester M13 9PL, UK
   \and
   Canadian Institute for Theoretical Astrophysics, 60 St George St, University of Toronto, Toronto, ON M5S 3H8, Canada
   \and
   Department of Physics, 60 St George St, University of Toronto, Toronto, ON M5S 3H8, Canada
   \and
   Department of Astronomy, University of Geneva, Chemin Pegasi 51, 1290 Sauverny, Switzerland
   \and
   Instituto de F\'{i}sica de Cantabria (CSIC-UC), Avenida. Lastros s/n. 39005 Santander, Spain
   \and
   IPAC, California Institute of Technology, 1200 E. California Blvd. Pasadena, CA 91125, USA
   \and
   Department of Astronomy/Steward Observatory, University of Arizona, 933 N. Cherry Avenue, Tucson, AZ 85721, USA
   \and
   Department of Astrophysical Sciences, Princeton University, Princeton, NJ 08544, USA
   \and
   Institute for Cosmic Ray Research, The University of Tokyo, 5-1-5 Kashiwanoha, Kashiwa, Chiba 277-8582, Japan
   \and
   Kavli Institute for Astronomy and Astrophysics, Peking University, Beijing 100871, China
   \and
   Observat\'orio Nacional, Rua General Jos\'e Cristino, 77 - Bairro Imperial de S\~ao Crist\'ov\~ao, Rio de Janeiro, 20921-400, Brazil
   \and
   Department of Space, Earth and Environment, Chalmers University of Technology, SE-412 96 Gothenburg, Sweden
   \and
   Centre for Astrophysics and Supercomputing, Swinburne University of Technology, Melbourne, VIC 3122, Australia
   \and
   Institute of Science and Technology Austria (ISTA), Am Campus 1, 3400 Klosterneuburg, Austria
   \and
   INAF -- OAS, Osservatorio di Astrofisica e Scienza dello Spazio di Bologna, via Gobetti 93/3, I-40129 Bologna, Italy
   \and
   MIT Kavli Institute for Astrophysics and Space Research, 70 Vassar Street, Cambridge, MA 02139, USA
   \and
   Department of Physics, Graduate School of Science, The University of Tokyo, 7-3-1 Hongo, Bunkyo, Tokyo 113-0033, Japan
   \and
   Department of Physics \& Astronomy, Tufts University, Medford, MA 02155, USA
   \and
   Department of Physics and Astronomy, Texas A\& M University, College Station, TX, 77843-4242 USA
   \and
   George P. and Cynthia Woods Mitchell Institute for Fundamental Physics and Astronomy, Texas A\& M University, College Station, TX, 77843-4242 USA
   \and
   CNRS, Centre de Recherche Astrophysique de Lyon UMR 5574, Universit\'{e} Lyon 1, ENS de Lyon, F-69230 Saint-Genis Laval, France
   \and
   Department of Astronomy, University of Maryland, College Park, 20742, USA
   \and
   Department of Astronomy, University of California, Berkeley, Berkeley, CA 94720, USA
   \and
   Center for Astrophysics $|$ Harvard \& Smithsonian, 60 Garden St., Cambridge, MA 02138, USA
   \and
   Department of Physics and Astronomy, University of California, Los Angeles, CA, 90095, USA
   \and
   INAF -- Osservatorio Astronomico di Roma, Via Frascati 33, I-00078 Monte Porzio Catone, Italy
   \and
   NRC Herzberg, 5071 West Saanich Rd, Victoria, BC V9E 2E7, Canada
   \and
   School of Earth and Space Exploration, Arizona State University, Tempe, AZ 85287-6004, USA
   }

  \date{Received XX XX 2025; accepted XX XX 2025}
 
  \abstract{We report the discovery of a doubly-imaged Little Red Dot (LRD) candidate behind the galaxy cluster Abell 383, which we dub ֿA383-LRD1. Initially classified as a dropout galaxy in HST imaging with several ground-based emission line detections placing it at $z_{\mathrm{spec}}=6.027$, new JWST/NIRCam observations taken as part of the cycle 4 VENUS survey now reveal that the source consists of two underlying components: A red point-source with a V-shaped SED consistent with LRD selection criteria, and a nearby ($\sim 380$\,pc) compact blue companion which was the main contributor to the previous rest-frame UV detections. Based on lensing symmetry and its SED, the LRD appears to lie at a similar redshift as well. The magnification of the two images of A383-LRD1 is $\mu_{\mathrm{A}}=16.2\pm1.2$ and $\mu_\mathrm{B}=9.0\pm0.6$, respectively, and the predicted time delay between them is $\Delta t_{\mathrm{grav}}=5.20\pm0.14$\,yr ($\sim0.7$\,yr in the rest-frame). After correcting for the lensing magnification, we derive an absolute magnitude of $M_{\mathrm{UV,LRD}}=-16.8\pm 0.3$ for the LRD, and $M_{\mathrm{UV,BC}}=-18.2\pm 0.2$ for the blue companion.  We perform SED fits to both components, revealing the LRD to be best fitted with a black hole star (BH*) model and a substantial host galaxy, and the blue companion with an extremely young, emission-line dominated star-forming nebula. A383-LRD1 represents the second known multiply-imaged LRD detected to date, following A2744-QSO1, and to our knowledge, the first LRD system with a confirmed detection of [C\,\textsc{ii}]$\lambda158 \ \mu$m emission from ALMA observations. Thanks to lensing magnification, this system opens a unique door to study the relation between a LRD, its host galaxy, and its environment, and represents a prime candidate for deep JWST spectroscopy and high-resolution ALMA follow-up observations.}

   \keywords{Quasars: emission lines -- Galaxies: high-redshift -- Gravitational lensing: strong -- Quasars: individual: A383-LRD1 -- Quasars: supermassive black holes}

   \titlerunning{A new multiply imaged LRD: A383-LRD1}
   \authorrunning{M. Golubchik et al.}

   \maketitle
%
\section{Introduction}\label{sec:intro}
Little Red Dots (LRDs) represent a new class of compact and extremely red, mostly high-redshift ($z\gtrsim4$) sources discovered with the JWST \citep[e.g.,][]{kocevski23,harikane23b,furtak23d,matthee24a,greene24,labbe25}. Many hundreds of LRD candidates have been detected to date \citep[e.g.,][]{kokorev24a,labbe25,Kocevski2025}, out to $z\sim9-10$ \citep[e.g.,][]{Taylor2025,Tanaka2025z10}, and with some analogs in the local Universe \citep[e.g.,][]{Lin2025,Ji2025b}. LRDs show unique spectral features such as a V-shaped SED \citep[e.g.,][]{setton24,labbe25} and have broad emission lines \citep[e.g.,][]{kocevski23,matthee24a,furtak24b,greene24} typical of type~1 active galactic nuclei (AGN), but they lack significant X-ray \citep[e.g.,][]{Lambrides24,yue24,maiolino25} and radio \citep[e.g.,][]{gloudemans25} emission. Initially thought to be strongly dust-obscured AGN \citep[e.g.,][]{furtak24b,greene24}, lack of infrared (IR) dust emission \citep[e.g.,][]{akins24,setton25,Leung2025} and the discovery in higher resolution spectra of significant absorption near the broad Balmer lines \citep[e.g.,][]{ji25,deugenio25} of LRDs, as well as extreme Balmer-breaks that could not be explained with stars \citep[e.g.,][]{naidu25,degraaff25}, suggest that LRDs consist of an AGN enshrouded in a dense and possibly turbulent shell of Hydrogen gas \citep[e.g.,][]{inayoshi25,madau25,pacucci24}; often referred to as the `Black Hole Star' \citep[BH*; e.g.,][]{naidu25} model. At the time of writing, this BH* model with varying degrees of host-galaxy contributions likely represents the most commonly accepted model for LRDs \cite[e.g.][]{deGraaff2025arXiv251121820D}.

Gravitational lensing has played a critical role in studying LRDs. One of the first archetypes for a LRD is A2744-QSO1 at $z=7.04$ \citep{furtak23d,furtak24b}, lensed and multiply imaged by the galaxy cluster Abell~2744. Thanks to the lensing magnification, A2744-QSO1 led to the most stringent upper limit on the size of a LRD to date ($r<<30$\,pc) \citep{furtak23d}, which cemented the AGN nature of LRDs early on and suggested the complete absence of a host-galaxy in this object \citep[e.g.,][]{furtak24b,ji25,Juodzbalis2025}. With the often decades-long arrival-time delays between multiple images, it also becomes possible to probe the variability of AGN over long time scales in a much shorter monitoring time \citep[e.g.,][]{Williams2021RMlensedQSO,golubchik24}, which led, for example, to the detection of spectroscopic variability in A2744-QSO1 \citep[][]{ji25,Furtak2025}. More recently, \citet{Juodzbalis2025} made an attempt to dynamically measure the black hole mass in A2744-QSO1, an unprecedented feat at such a high redshift ($z=7$), only made possible with JWST and by strong-lensing (SL) magnification.

So far, however, the sample of lensed LRDs is limited to only a few \citep[e.g.,][]{kokorev23,greene24,killi24,tripodi24}, and only one of those is multiply-imaged, A2744-QSO1 \citet{furtak24b}, although a couple other candidates are reported in Zhang et al. (\emph{in preparation}).
Here, we report the detection of another multiply-imaged LRD candidate\footnote{Note that while, as we show here, the object passes the LRD photometric selection criteria, we nominally refer to it as a candidate until spectroscopy becomes available.} - dubbed hereafter A383-LRD1, located behind the well-known SL cluster Abell~383 \citep[$z_\mathrm{d}=0.187$; A383 hereafter;][]{abell89}. A383-LRD1 curiously has a close-by blue companion, which is also compact. The latter is particularly intriguing since a significant portion of LRDs have now been observed to have nearby UV emission, or blue companions \citep[e.g.,][]{Rinaldi2025,naidu25}, but their nature remains uncertain. For example, \citet{Chen2025} suggest that this blue emission in the vicinity of LRDs could be arising from low-density, metal-poor gas that is photo-ionized by the AGN. This might also, for example, explain the presence of narrow high ionization UV lines detected in some LRDs \citep{Tang2025}. With the system presented in this work, we could now investigate this, and other LRD with blue companion properties, with the aid of SL.

This paper is organized as follows: The data are described in section~\ref{sec:data}. In section~\ref{sec:analysis}, we present the object, and the photometric and SL analysis. In section~\ref{sec:results}, we discuss our results for both the LRD candidate and its blue companion, and in section~\ref{sec:summary} we give a brief summary and conclusion. Throughout this work, we use a flat $\Lambda$CDM cosmology with $H_0=70$ km s$^{-1}$ Mpc$^{-1}$, $\Omega_{\Lambda}=0.7$, and $\Omega_\mathrm{m}=0.3$, and use AB magnitudes \citep{oke83}. Errors are typically $1\sigma$ unless noted otherwise.

\begin{figure*}
    \centering
    \includegraphics[width=\textwidth]{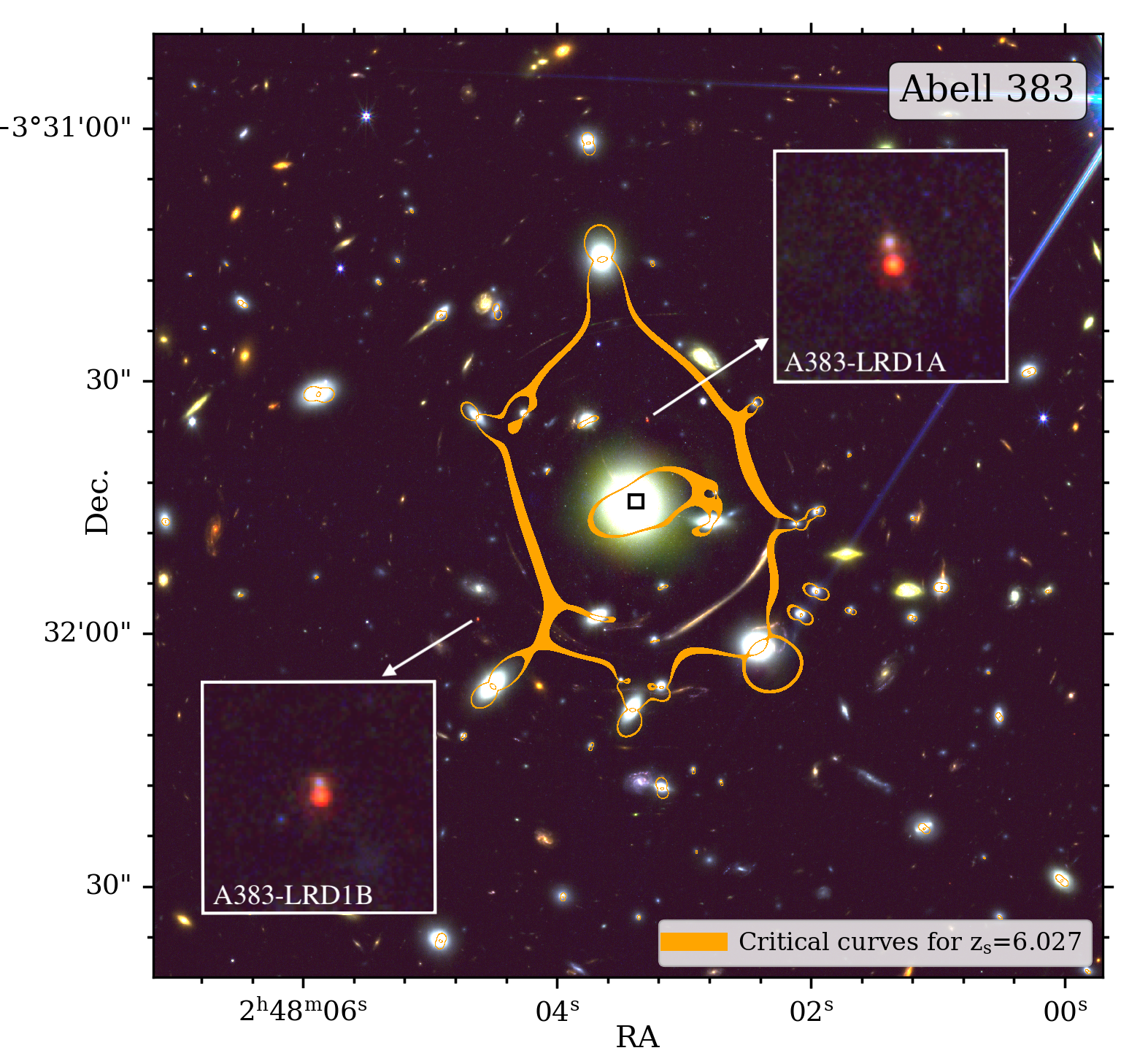}
    \caption{Abell383-LRD1 -- a new multiply-imaged LRD candidate. The critical curves for $z_{\mathrm{s}}=6.027$ from our new SL model (see section~\ref{lensing}) are shown in orange, over a color-composite image of A383 (${z_{\mathrm{d}}=0.187}$), constructed using the JWST/NIRCam imaging from the VENUS survey (Blue: F090W, Green: F277W, Red: F444W). The predicted position of the third de-magnified image of the system is marked with a black box. The third image is obscured by the BCG and predicted to be too demagnified to be observed. The inlets correspond to $3\arcsec\times3\arcsec$ cutouts and show the two multiple images of A383-LRD1, with the blue companion clearly seen in both images.} 
    \label{fig:LRD}
\end{figure*}

\section{Data} \label{sec:data}
The SL cluster A383 was recently observed with the \textit{Near Infrared Camera} \citep[NIRCam;][]{rieke23} aboard the JWST \citep[][]{gardner23,mcelwain23} as a part of the \textit{Vast Exploration for Nascent, Unexplored Sources} (VENUS) survey (Program ID: GO-6882; PIs: S.~Fujimoto \& D. Coe) in September~2025. NIRCam imaging was obtained in 10 filters (F090W, F115W, F150W, F200W, F210M, F277W, F300W, F356W, F410M, and F444W), with exposure times spanning $0.7-3.7$\,h. The observations achieve $5\sigma$ depths of 28\,magnitudes in all bands and were reduced and drizzled into mosaics with the \texttt{Grism redshift \& line analysis software for space-based slitless spectroscopy} \citep[\texttt{grizli};][]{grizli23}. We refer the reader to Fujimoto et al.\ (in prep.) and Kokorev et al.\ (in prep.) for more details on the VENUS observations, data reduction, and catalogs.

To complement the JWST observations, we use optical and ultra-violet (UV) imaging data from the \textit{Hubble Space Telescope} (HST), obtained as part of the \textit{The Cluster Lensing and Supernova Survey with Hubble} program \citep[CLASH; Program ID: GO-12065;][]{postman12}. These ancillary data include six broad-band filters of the \textit{Advanced Camera for Surveys} (ACS): F435W, F475W, F606W, F625W, F775W, and F814W. Note WFC3 images from CLASH exist as well but are not used here. The above HST imaging was re-processed as part of the \textit{Complete Hubble Archive for Galaxy Evolution} (CHArGE) initiative, which performed uniform processing of all archival HST imaging \citep{Kokorev2022}, and was drizzled into mosaics matching the VENUS JWST images with \texttt{grizli}.

Additional ancillary data include ground-based spectroscopy with the \textit{DEep Imaging Multi-Object Spectrograph} (DEIMOS) on \textit{Keck}, published in \citet{Richard2011}, as well as XSHOOTER, published in \citet{Stark2015}, and \textit{Multi Unit Spectroscopic Explorer} \citep[MUSE;][]{bacon10}, on ESO's \textit{Very Large Telescope} (VLT). Finally, our LRD was also detected with the \textit{Atacama Large Millimeter/sub-millimeter Array} (ALMA) in \citet{Knudsen2016}.

\section{The Object and Analysis} \label{sec:analysis}

\begin{figure*}
    \centering
    \includegraphics[width=\textwidth]{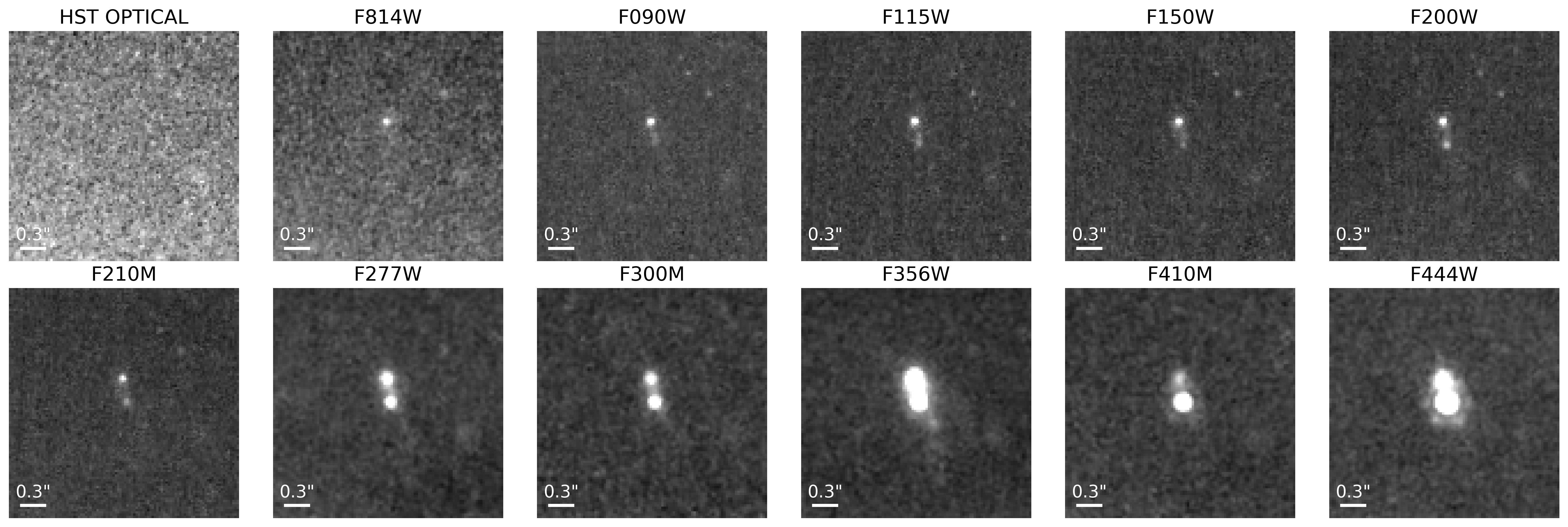}
    \caption{Cutouts of image A of A383–LRD1 in the JWST/NIRCam bands from the VENUS program and in the available HST bands. The upper–left panel shows a stacked image from the ACS bands-F435W+F475W+F606W+F625W+F775W, where the source remains undetected. Each cutout covers a $3 \arcsec \times 3\arcsec$ field of view. Only the blue companion is visible in the F814W image, while the LRD becomes more prominently detectable towards redder wavelengths. The blue companion has a noticeable flux excesses in F356W and F444W, likely due to strong H$\beta$+[O\,\textsc{iii}] and H$\alpha$ emission lines, respectively, at $z=6.027$.} 
    \label{fig:stamps}
\end{figure*}

The JWST/NIRCam imaging (see Fig~\ref{fig:LRD}) reveals a notable doubly-imaged source, which consists of two distinct components: A red and compact source which as we show here is likely a LRD, and a compact blue companion. Figure \ref{fig:stamps} illustrates the source in all available bands. It can be seen that both A383-LRD1 and the blue companion are undetected in filters bluer than F814W, and the LRD component becomes increasingly more prominent towards the redder bands. We also note that some additional faint emission appears in the F356W band just below the LRD; and there also seems to be faint bridge extending between the LRD and blue companion, which is noticeable in various bands (see Fig.\ref{fig:stamps}). This system was in fact, previously known as a multiply-imaged high-redshift source from the HST CLASH observations (system 5 in \citealt{Richard2011,Zitrin2011}), and was also spectroscopically confirmed at $z_{\mathrm{spec}}=6.027$ with its Lyman-$\alpha$ (Ly$\alpha$) line \citep{Richard2011}. \citet{Stark2015} then measured its extreme Ly$\alpha$ equivalent width (EW) of $\mathrm{EW}_0=138$\,\AA, and detected C\,\textsc{iii}]$\lambda1909$\AA\ emission, and \citet{Knudsen2016} detected [C\,\textsc{ii}]$\lambda158 \ \mu$m from it. In these previous works, the galaxy was classified as `old' due to a flux excess measured with \textit{Spitzer}/IRAC, interpreted as a Balmer-break \citep{Stark2015}. Now, with the superior spatial resolution of the JWST, we can clearly see that the rest-frame optical emission originates from a separate component, which is the LRD (see Fig.~\ref{fig:LRD}). In contrast, given that the LRD is barely detected in the bluer bands (see Fig.~\ref{fig:stamps}), the blue companion is the component that was identified in the HST images and thus is also likely the origin of the UV emission lines. Moreover, if Ly$\alpha$ were originated from the LRD, it would have been detected in the F814W band to the same level as the blue companion, but this is not observed.

Based on the gravitational lensing symmetry, and the spectral energy distributions (SED; Fig. \ref{fig:SED galaxy fits}), both components are consistent with lying at a similar redshift. While only future integral field unit (IFU) observations could spatially disentangle the two components, and verify their physical relation, throughout the rest of this work we assume that both lie at the same redshift of $z=6.027$. The properties of both images are summarized in Tab.~\ref{tab:fluxes_all_combined}. We conduct a thorough photometric analysis of both components in section~\ref{phot}, derive the lensing properties in section~\ref{lensing}, measure sizes in section~\ref{size}, and finally perform a SED modeling analysis in section~\ref{sed}.

\subsection{Photometry and color-analysis} \label{phot}

\begin{table*}
\centering
\caption{Photometric and lensing properties for A383-LRD1 and its blue companion.}
\label{tab:fluxes_all_combined}
\small

\begin{tabularx}{\textwidth}{@{}l*{8}{>{\centering\arraybackslash}X}@{}}
\toprule
ID & R.A.~[Deg.] & Dec.~[Deg.] & $\Delta t_{\mathrm{grav}}~[\mathrm{yr}]$ & $\mu$ & $\mu_t$ & $\mu_r$ & F435W & F475W \\
\midrule
A383-LRD1A & 42.013686 & -3.526365 & 0 & $16.2\pm1.2$ & $3.7\pm0.1$ & $4.2\pm0.1$ & $-2.2 \pm 12.4$ & $-1.5 \pm 11.7$ \\
Blue Companion-A & 42.013702 & -3.526283 & 0 & $16.2\pm1.2$ & $3.8\pm0.1$& $4.1\pm0.1$& $2.2 \pm 12.4$ & $-0.6 \pm 11.7$ \\
A383-LRD1B & 42.019241 & -3.532926 & $-5.2\pm0.1$ & $9.0\pm0.6$ & $4.8\pm0.2$ & $1.9\pm0.1$& $5.8 \pm 12.8$ & $4.4 \pm 12.4$ \\
Blue Companion-B & 42.019246 & -3.532876 & $-5.1\pm0.1$ & $9.0\pm0.6$ & $4.8\pm0.2$& $1.9\pm0.1$ &  $-0.5 \pm 12.8$ & $-2.7 \pm 12.2$ \\
\bottomrule
\end{tabularx}

\vspace{4pt}

\begin{tabularx}{\textwidth}{*{7}{>{\centering\arraybackslash}X}}
\toprule
F606W & F625W & F775W & F814W & F090W & F115W & F150W \\
\midrule
$1.3 \pm 5.7$ & $-11.9 \pm 15.1$ & $4.6 \pm 17.2$ & $18.0 \pm 10.4$ & $36.8 \pm 7.7$ & $68.6 \pm 8.6$ & $58.2 \pm 9.4$ \\
$4.1 \pm 5.7$ & $-7.9 \pm 14.8$ & $38.6 \pm 17.4$ & $88.2 \pm 10.5$ & $188.1 \pm 8.2$ & $247.0 \pm 9.2$ & $231.4 \pm 9.9$ \\
$-2.3 \pm 5.5$ & $18.1 \pm 14.6$ & $14.9 \pm 16.1$ & $17.4 \pm 9.3$ & $20.5 \pm 4.5$ & $32.2 \pm 4.6$ & $45.2 \pm 4.8$ \\
$-1.8 \pm 5.5$ & $2.1 \pm 13.8$ & $3.9 \pm 16.1$ & $67.2 \pm 9.5$ & $111.1 \pm 4.7$ & $145.1 \pm 4.9$ & $145.6 \pm 5.1$ \\
\bottomrule
\end{tabularx}

\vspace{4pt}

\begin{tabularx}{\textwidth}{*{7}{>{\centering\arraybackslash}X}}
\toprule
F200W & F210M & F277W & F300M & F356W & F410M & F444W \\
\midrule
$116.8 \pm 9.0$ & $119.8 \pm 11.8$ & $201.5 \pm 5.8$ & $277.4 \pm 7.6$ & $608.4 \pm 6.4$ & $649.8 \pm 8.5$ & $1311.4 \pm 7.9$ \\
$219.3 \pm 9.9$ & $209.2 \pm 12.0$ & $213.2 \pm 5.9$ & $189.1 \pm 7.6$ & $559.4 \pm 6.4$ & $122.6 \pm 7.1$ & $261.9 \pm 5.7$ \\
$56.2 \pm 4.8$ & $50.6 \pm 5.1$ & $113.5 \pm 4.1$ & $160.0 \pm 5.6$ & $395.6 \pm 5.5$ & $389.7 \pm 7.5$ & $830.3 \pm 7.2$ \\
$127.6 \pm 5.1$ & $124.5 \pm 5.2$ & $127.7 \pm 4.1$ & $128.3 \pm 5.3$ & $375.9 \pm 5.8$ & $95.1 \pm 5.9$ & $218.5 \pm 5.3$ \\
\bottomrule
\end{tabularx}

\vspace{4pt}

\footnotesize The lensing magnification, time delays, and photometric flux densities (nJy) and $1\sigma$ errors for both the LRD and blue companion in the two images of A383-QSO1 (see sec.\ref{phot} for more details). \emph{Column 1:}  ID \emph{Columns 2-3:} R.A., Dec. \emph{Columns 4-7:} gravitational time delays, magnifications, and their tangential and radial components. \emph{Columns 8-13:} The fluxes measured in available HST bands, and \emph{Columns 14-23:} The fluxes measured in available JWST bands. Note: Columns 10-23 wrap to subsequent lines below Columns 1-9.

\end{table*}

We measure simple aperture photometry of both A383-LRD1 and the blue companion, in both images, in all the available JWST bands and the HST filters from section \ref{sec:data}, with the \texttt{photutils} package \citep[\texttt{v2.2.0};][]{Bradley25}. The apertures, of 0.2\arcsec\ and 0.16\arcsec\ diameter in images A and B respectively, are carefully chosen to limit contamination by either companion. The photometry is then measured on a background-subtracted image and corrected for aperture losses due to the point-spread-function (PSF) size in each band. We estimate the errors by placing empty apertures in the vicinity of the source to measure the local background contribution. This approach was shown to perform well for point-sources in crowded lensed fields (see e.g.\ Furtak et al.\ in prep.). The resulting flux densities are listed in Tab.~\ref{tab:fluxes_all_combined}, and can be seen in Fig.~\ref{fig:SED galaxy fits}. We use the F115W band, corresponding to the rest-frame 1500\,\AA\ range, to measure the rest-frame UV luminosity, obtaining $M_{\mathrm{UV,LRD-A}}=-16.8\pm 0.3$, $M_{\mathrm{UV,LRD-B}}=-16.7\pm 0.3$  for the LRD, and $M_{\mathrm{UV,BC-A}}=-18.2\pm 0.2$, $M_{\mathrm{UV,BC-B}}=-18.3\pm 0.2$ for the blue companion in both images after de-magnification (see Tab.~\ref{tab:fluxes_all_combined} and section~\ref{lensing}). The values are consistent between the two images, with no sign of variability.

The measured colors of the LRD in image A (i.e., the brightest of the two) are:
\[
\mathrm{F150W}-\mathrm{F200W} = 0.75 \pm 0.26
\]
\[
\mathrm{F277W}-\mathrm{F356W} = 1.20 \pm 0.05
\]
\[
\mathrm{F277W}-\mathrm{F444W} = 2.03 \pm 0.04
\]
We use the measured colors to verify whether this component of A383-QSO1 satisfies established LRD selection criteria. Following the approach of \citet{kokorev24a}, we first examine their `red2' criteria defined for LRDs at $z>6$ (for more detalis, see their section~3.1):
\begin{gather*} 
\mathtt{red \ 2} = \mathrm{F150W}-\mathrm{F200W} < 0.8 \ \& 
~\mathrm{F277W}-\mathrm{F356W} > 0.6 \ \& \\
\mathrm{F277W}-\mathrm{F444W} > 0.7
\end{gather*} 
The measured colors of A383-LRD1 satisfy the `red2' selection criteria.
We then evaluate the object’s compactness by measuring the ratio between the F444W flux in a $0.4 \arcsec$ and a $0.2 \arcsec$ apertures. Even though the $0.4 \arcsec$ measurement includes some flux from the blue companion, the LRD still satisfies the compactness criteria. In addition, we verify that its colors are inconsistent with colors expected for brown dwarfs. For completeness, we repeat the same analysis also for the criteria shown in \citet{labbe25}, \citet{greene24}, \citet{akins24}, \citet{Tanaka2024} and \citet{degraaff25}, all suggest it is indeed a LRD. To further assess consistency, we also compute the $\beta_{\mathrm{F277W}-\mathrm{F356W}}$ and $\beta_{\mathrm{F277W}-\mathrm{F410M}}$ slopes following \citet{Kocevski2025}, which are used to exclude sources dominated by strong optical line emission (see their section~3.1). Although the object shows enhanced F356W and F444W fluxes relative to F410M, which respectively correspond to the H$\beta$+[O\,\textsc{iii}] and H$\alpha$ emission lines at $z=6.027$, the calculated $\beta_{\mathrm{F277W}-\mathrm{F356W}}$ and $\beta_{\mathrm{F277W}-\mathrm{F410M}}$ slopes are consistent with LRDs, rather than emission line objects. To further make sure, we manually adjust these flux excesses to match the underlying continuum (interpolated or extrapolated from nearby bands), which still yields colors consistent with the LRD selection criteria. Finally, if we combine the photometry of the LRD and the blue companion together -- for example, to imitate a configuration in which a LRD and its blue companion are unresolved (i.e. unlensed) -- the combined object would still pass the more stringent 'red2' selection criteria. 

\subsection{Strong lensing mass modeling} \label{lensing}
Our new lens model builds on previous models for the cluster \citep{Richard2011,zitrin15a}, and is constructed here using a revised version of the parametric method by  \citet[][for more details see also \citealt{pascale22a,furtak23c}]{zitrin15a}. In the model, cluster member galaxies are represented by double Pseudo-Isothermal Ellipsoids (dPIE; \citealt{eliasdottir07}), and the large-scale dark matter distribution is described by a diffuse halo modeled as a Pseudo-Isothermal Elliptical Mass Distribution (PIEMD; e.g.\ \citealt{keeton01a}). 
We use 14 multiple-image systems as constraints, 6 confirmed with spectroscopy for at least one image (described in \citealt{zitrin15a}).

Cluster members are chosen matching the red sequences formed in the NIRCam/F090W and F150W; ACS/F814W and NIRCam/F150W; and ACS/F606W and F814W spaces, together with data from MUSE. The model is minimized using a positional uncertainty of $0.1\arcsec$ for all images and $0.05\arcsec$ for the southern giant arc. All multiple images are reproduced by the model with a $\Delta_{\mathrm{RMS}}$ between model and observations of $0.36\arcsec$. Uncertainties are obtained by running a separate MC chain with a positional uncertainty of $0.5\arcsec$. With this the best-fit model yields a reduced chi-square of $\simeq\chi^{2}/\mathrm{DOF}=23.9/20\simeq1.2$.

We infer the magnification for both images [given in the following format: best(average)$\pm$ uncertainty]: $\mu_A = 16.2(17.9)\pm1.2$, for image A (5.1 in \citealt{Richard2011}) and $\mu_B = 9.0(10.0)\pm0.6$ for image B (5.2 in \citealt{Richard2011}). We note however that an additional systematic error of $\sim20\%$ could be assumed. Also note, previous models have inferred magnification of $\approx11$ and $\approx6$ for the two images, respectively \citep{Richard2011}. 
A third, radial, de-magnified ($\mu \approx 0.2$) image is predicted near the BCG, but its location (marked on Fig. \ref{fig:LRD}), and expected flux ($\sim5$ magnitudes fainter than image A), render it unobservable. 
The time delays between the two observed images according to our model corresponds to image A arriving $5.2(4.9)\pm0.1$\,yr after image B (we estimate a systematic uncertainty of $\sim0.5$\,yr from examining a range of models).
The blue companion in image A is resolved through lensing, $0.3\arcsec$ away from the LRD, which  -- from the above magnifications -- corresponds to $\sim400$\,pc in the source plane assuming $z=6.027$. We then de-lens both images to the source plane, and find the distance between the centers of the LRD and blue companion to be $\sim380$\,pc, in agreement with the above. As only the redshift of the blue companion is spectroscopically measured, we run an alternative lens model allowing the LRD redshift to be free. The model's predicted redshift for the LRD agrees with the redshift of the blue companion, $z=6.027$, within 1$\sigma$.

\subsection{Size measurements} \label{size}
Both the LRD and the blue companion appear as compact sources in the color image. We note that the compactness of the LRD candidate (discussed above in section \ref{phot} and defined using the ratio between the F444W flux in a $0.4 \arcsec$ and a $0.2 \arcsec$ apertures) is already sufficient -- together with its colors -- to classify it photometrically as a LRD. However, given the LRD is lensed and that it has a compact blue companion, it is interesting to further examine the relevant object sizes.

\begin{figure*}
    \includegraphics[width=0.49\textwidth]{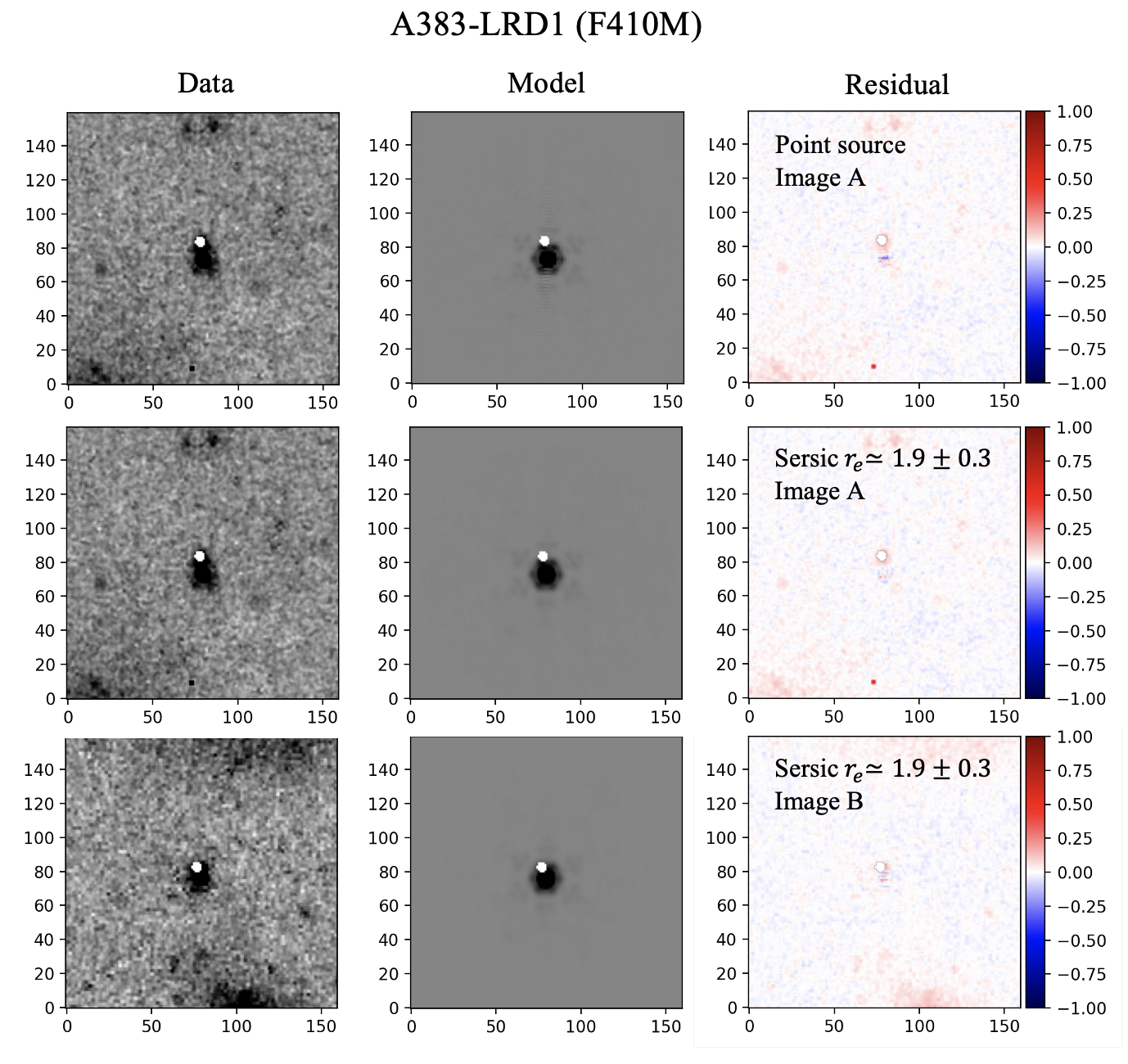}
    \includegraphics[width=0.49\textwidth]{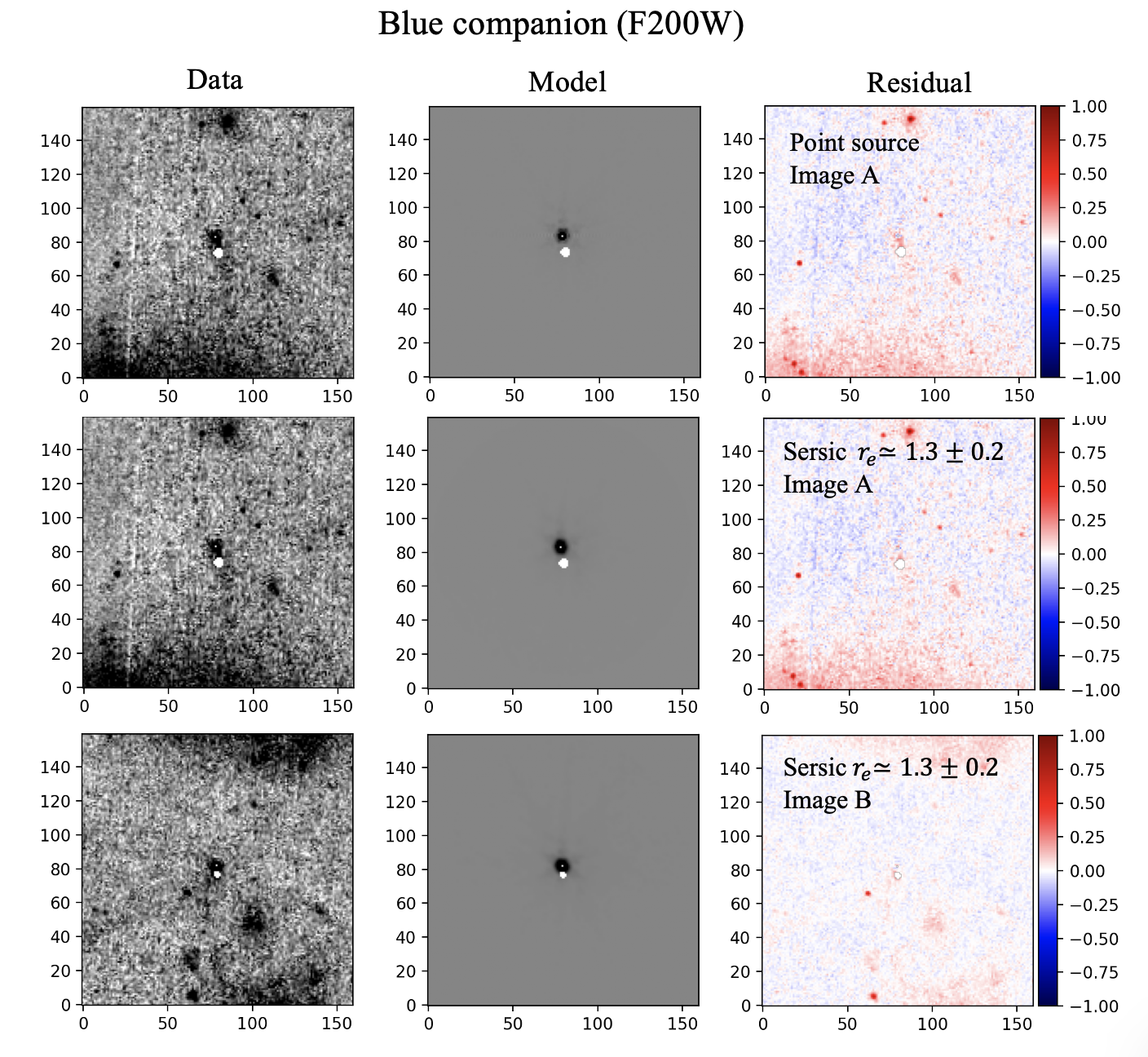}\\
    \includegraphics[width=0.49\textwidth]{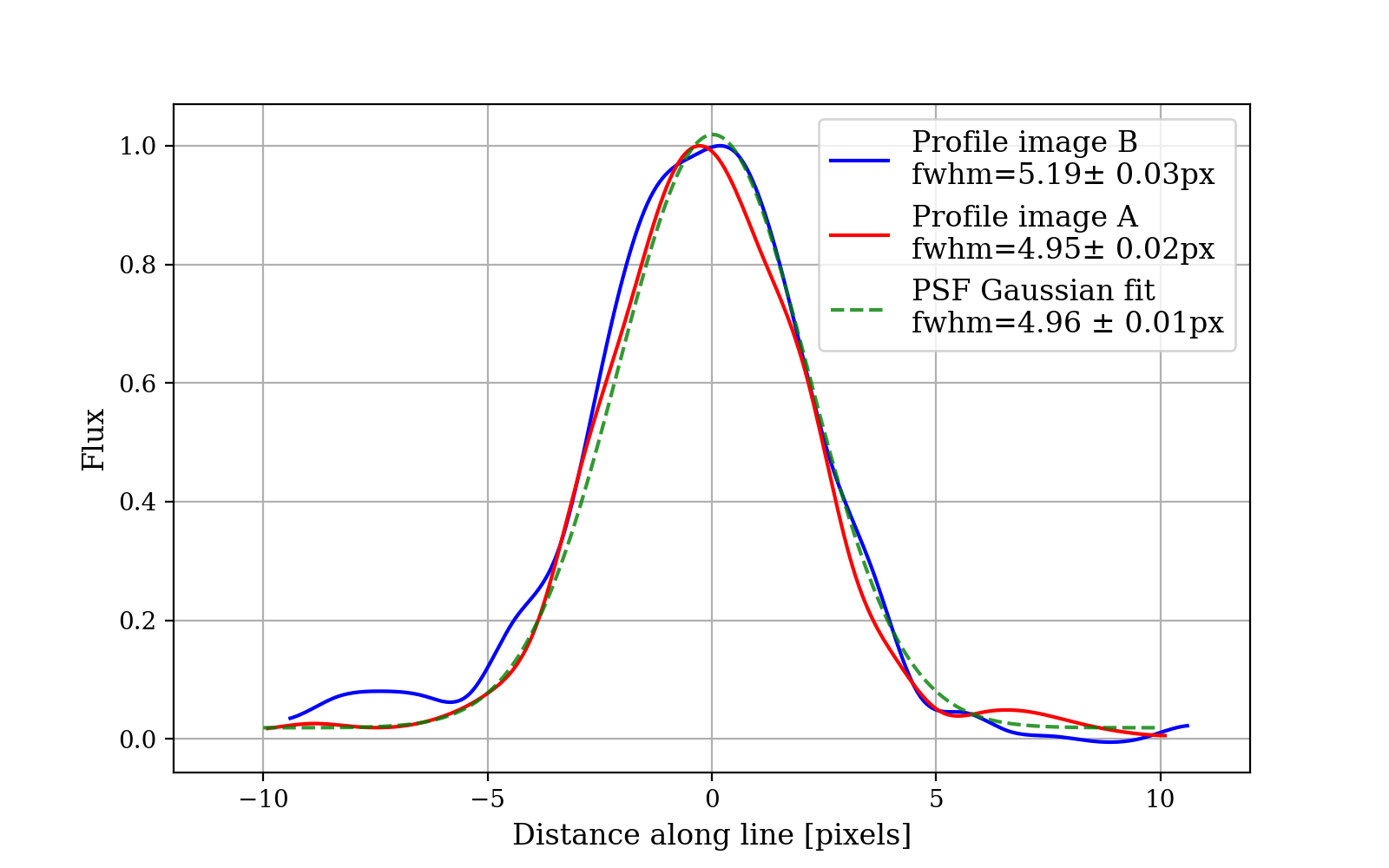}
    \includegraphics[width=0.49\textwidth]{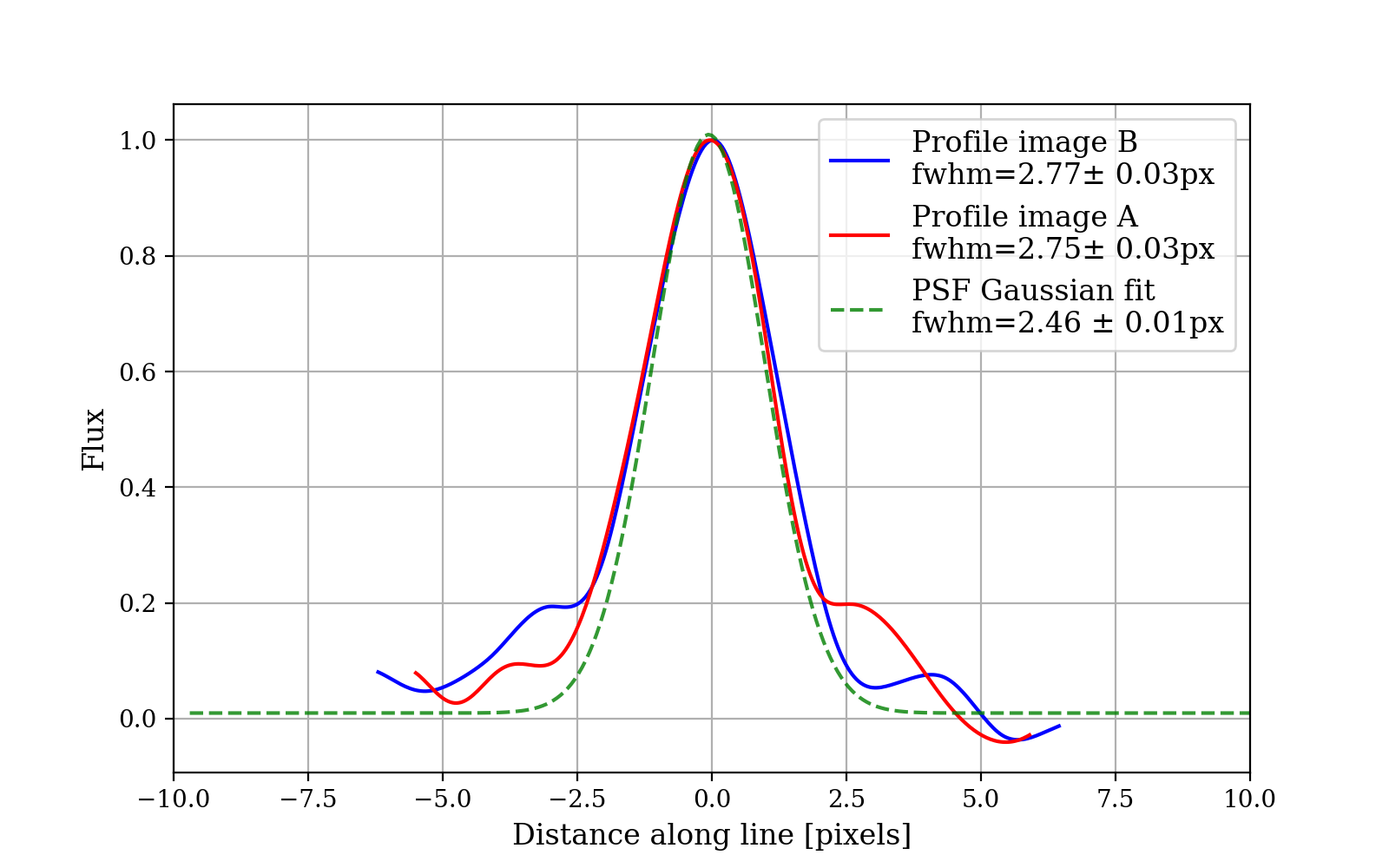}
    \caption{\texttt{pysersic} fits to the LRD (in the F410M band) and the blue companion (in the F200W band). We show a point-mass model fit for image A, and a Sersic profile fit for both images A and B. The small white region near the center indicates the masked component in each fit. For both the LRD and the blue companion, the two model types yield an excellent reconstruction, and the size (e.g., Sersic effective radius) remains the same between the two multiple images despite a factor of $\sim1.7$ in magnification ratio. This is also seen by simply comparing their model-independent light-profiles to each other and to the PSF profile, as seen in the bottom subfigures. See section \ref{size} for more details.} 
    \label{fig:size1}
\end{figure*}

To measure the sizes we analyze both multiple images with the \texttt{pysersic} program \citep{Pasha2023_pysersic}. We start with the LRD and use the F410M image -- where the blue companion is weakly seen, so its contribution can be easily masked but the LRD is prominently observed; and the F200W band, where the PSF is smaller and which we shall later use for constraining the physical size. For each band we employ a PSF image constructed from an ensemble of stars in the field. We manually mask the blue companion's pixels and run a fit to the LRD using a Sersic profile. Both images of the LRD (LRD1A and LRD1B) yield the same size of $r_{e}\simeq1.9\pm0.3$ pixels in the F410M band, and $r_{e}\simeq1.3\pm0.2$ pixels in the F200W band. We note however that the resulting parameter values are very sensitive to the choice of different mask regions and exact PSF used, and thus these should be taken with caution. We also try a \emph{point source} fit option which yields a similarly good reconstruction, although perhaps somewhat less clean visually (Fig. \ref{fig:size1}). The fact that both images of the LRD show the same size despite a ratio of $\sim1.7$ in magnifications and observed fluxes, suggest that the source is unresolved and consistent with a point source: Were it extended, the observed FWHM would be larger in the more magnified image. Note we also verified that the observed size remains similar regardless of the assumed model, by simply plotting the light profile of each of the LRD images and measuring the width. This as well as some examples for the fitting results are shown in Fig. \ref{fig:size1}. 

For measuring the size of the blue companion we do two things. First, we mask the LRD and repeat the fitting procedure for the blue companion over both images, in particular in bluer bands where it is the dominant light source and the LRD contribution is easily masked (see Fig. \ref{fig:stamps}). Then, we run a two-object Sersic profile fit for each of the images. We obtain for both multiple images of the blue companion a similar value of $r_{e}\simeq1.0\pm0.1$ to $r_{e}\simeq1.1\pm0.1$ in F090W, F115W, and F200W. The blue companion thus seems to also be consistent with a point source in those bands (Fig. \ref{fig:size1}). In other bands, however, such as the F277W or the F410M bands, we obtain that the effective radius of image A can be larger than that of image B, by factors of $\sim1.5-2.5$. This observed size ratio is consistent with the magnification ratio, implying that perhaps the blue companion, while compact, is only marginally (un)resolved and is thus not a point source in some of the redder bands. It is however unclear at present whether this is indeed the case or a result of the fitting procedure (given the strong contrast in F410M between LRD and the blue companion, for example, or the fact that there appear to be other structures involved which may bias the fit such as the faint bridge between the blue and red components seen in e.g. the F277W band). Another possibility is that the UV light in the blue companion is indeed coming from a smaller region than its optical light. We leave further investigation of this to future work.

We now translate the observed size measurements to constraints on the physical size of the sources. Assuming the LRD is indeed unresolved, we can adopt the nominal PSF FWHM of the F200W band ($0.064''$) as an upper limit on the source size. Accounting for both redshift and tangential magnification, we thus derive an upper limit on the intrinsic radius of $r_{\mathrm{intr}} \lessapprox 40\, \mathrm{pc}$ for the LRD. Note that while the PSF of the bluer bands would in principle lead to a stronger size constraint, we choose to use the F200W band because it is the bluest band in which the LRD is detected at above 10$\sigma$. In addition, despite higher magnification for A383-LRD1 compared to A2744-QSO1, the size constraint is slightly weaker, because the radial and tangential components of the magnification are relatively similar in our model.
For the blue companion, assuming it is only marginally (un)resolved, we measure a rough size of $r_{\mathrm{intr}} \approx 60\,\mathrm{pc}$, but note that its UV emitting part may be significantly smaller. If we assume the UV emission of the blue companion to be a point source, then -- adopting the F090W PSF -- it must be smaller than 20 parsec.

\subsection{SED modeling} \label{sed}
We perform SED modeling on the de-magnified photometry of both the LRD component, and the blue companion. The SED fitting is run on image A since it provides a higher signal-to-noise ratio (SNR), and the components appear more separated and thus the photometry (see section~\ref{phot}) is less likely to be contaminated. 

We first run a fit to both components with the the \texttt{Bayesian Analysis of GaLaxy sEds} tool \citep[\texttt{BEAGLE};][]{chevallard16} and SED templates by \citet{gutkin16}, which combine the latest \citet{bc03} stellar population synthesis models with nebular emission models from \texttt{Cloudy} \citep[][]{ferland13}. The redshift is fixed to the spectroscopic value \citep[$z_\mathrm{spec}=6.0265$;][]{Stark2015}. For the blue companion, we use the C\,\textsc{iii}]$\lambda\lambda1907,1909$\AA\ equivalent width (EW) of $22\pm7$\,\AA, measured by \citet{Stark2015}, as an additional constraint. We assume a delayed exponential star-formation history (SFH) with the possibility of an ongoing star-burst and a Small Magellanic Cloud (SMC) dust extinction law \citep[][]{pei92}, which has been shown to well match low-mass and low-metallicity galaxies at high redshift \citep[e.g.,][]{shivaei20}. Finally, we apply the \citet{inoue14} intergalactic medium attenuation models. Both fits have as free parameters the stellar mass $M_{\star}$, the star-formation rate (SFR) of the current star-burst, the maximum stellar age $t_{\mathrm{age}}$, the star-formation $e$-folding time-scale $\tau$, the dust extinction $\hat{\tau}_{\mathrm{V}}$, the metallicity $Z$, and the average galaxy-wide ionization parameter $\log(\hat{U})$. After some iteration (see discussion in section~\ref{sec:BC_discussion}), we lower the time over which the fitted SFR is computed to 1\,Myr instead of the 10\,Myr assumed by default in \texttt{BEAGLE}, in order to better fit the extreme [O\,\textsc{iii}] and H$\beta$ emission in the F356W band -- especially for the blue companion (see Fig.~\ref{fig:SED galaxy fits}).

While we achieve a decent fit to the LRD component with a dusty star-forming galaxy (DSFG) template, as further discussed in section~\ref{sec:LRD_discussion}, the standard \texttt{BEAGLE} templates do not include LRD-specific models such as e.g.\ a BH* component. We therefore run a second fit to the LRD component only, using the \texttt{Bayesian Analysis of Galaxies for Physical Inference and Parameter EStimation} code \citep[\texttt{bagpipes};][]{Carnall2018}. This version of the code was modified by \citet{Taylor2025} to fit a composite model of a star-forming host-galaxy with a BH*-type AGN. The latter is included as an interpolation over a set of \texttt{CLOUDY} models designed to imitate the BH* dense hydrogen gas shell emission \citep[e.g.,][]{naidu25}, with a luminosity $L_{5100}$ as a normalization parameter. We refer the reader to \citet{Taylor2025} for the full details on the models and the fitting procedure. Note that the BH* model constrains the gas properties, and does not include the BH mass. The free parameters in particular are listed in their Tab.~2. It should be emphasized that in the absence of rest-frame optical spectroscopy of the LRD component, many of the BH* model's gas (or \texttt{CLOUDY} grid) parameters remain largely underconstrained. Our photometry does however probe the Balmer-break and rest-frame optical continuum, which means that we can, under the assumed model at least, constrain the gas density $n_{\mathrm{H}}$, the column density $N_{\mathrm{H}}$, and the turbulence velocity $v_{\mathrm{turb}}$ \citep[e.g.,][]{ji25}.

\section{Results and discussion} \label{sec:results}

\begin{figure*}
    \centering
    \includegraphics[width=\textwidth]{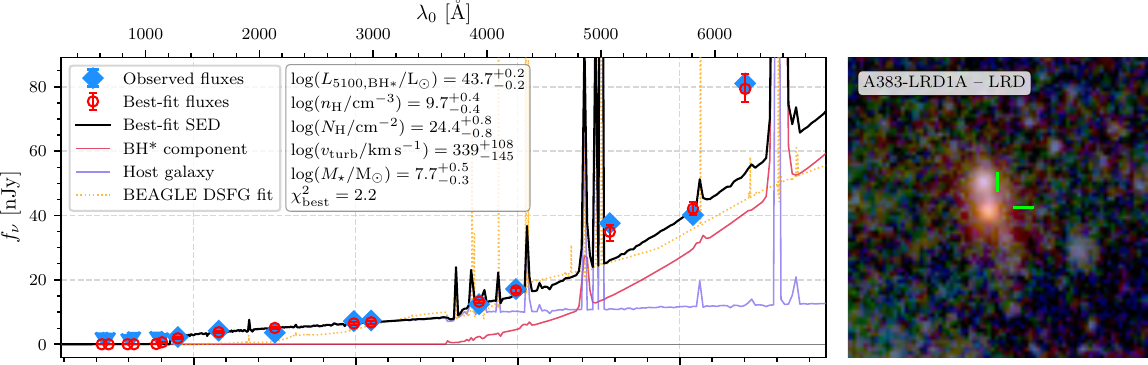}
    \includegraphics[width=\textwidth]{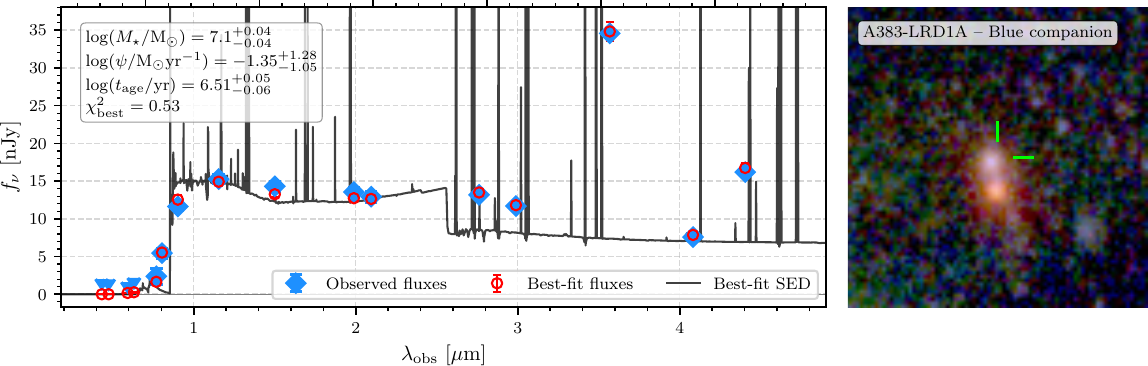}
    \caption{Photometry and SED-modeling of A383-LRD1 image~A, with composite-color image cutouts on the right. The de-magnified observed fluxes are shown in blue, with $2\sigma$ upper limits in the case of a non-detection. The best-fit (maximum-a-posteriori) fluxes are shown in red, and the best-fit SED in black. \textit{Top}: \texttt{Bagpipes} LRD fit to the LRD component. The LRD SED model comprises a BH*-AGN component shown in red, and a host galaxy shown in purple. We also show the \texttt{BEAGLE} DSFG fit as the orange dotted line. \textit{Bottom:} \texttt{BEAGLE} star-forming galaxy fit to the blue companion.}
    \label{fig:SED galaxy fits}
\end{figure*}

The results of our analysis, the photometry, and the SED models, are presented in Fig.~\ref{fig:SED galaxy fits}. We find the red component to most likely be a LRD, as discussed below in section~\ref{sec:LRD_discussion}. The nature of the blue companion is discussed in section~\ref{sec:BC_discussion}.

\subsection{The LRD candidate} \label{sec:LRD_discussion}
We find A383-LRD1 to be best fit with the \texttt{bagpipes} BH* and a host galaxy model (see section~\ref{sed}), achieving a best-fit reduced $\chi^2=2.2$. The fit suggests the LRD to have a relatively small contribution from the BH* AGN ($\log(L_{5100}/\mathrm{L}_{\odot})=43.2\pm0.2$), and significant contribution from a host galaxy ($\log(M_{\star}/\mathrm{M}_{\odot})=7.7_{-0.3}^{+0.5}$), which is evolved enough to produce a small Balmer break ($\sim0.5$ magnitudes). At longer wavelengths than the break, the BH* continuum takes over and dominates the reddest bands, F356W, F410M, and F444W, as can be seen in the top panel of Fig.~\ref{fig:SED galaxy fits}. For the BH* model, we obtain a best-fit gas density of $\log(n_{\mathrm{H}}/\mathrm{cm}^{-3})=9.7\pm0.4$ and a turbulence velocity of $v_{\mathrm{turb}}=339_{-145}^{+108}\,\mathrm{km\,s}^{-1}$, which are similar to what was found for e.g.\ A2744-QSO1 \citep{ji25}, or the high-z LRD presented in \citet{Taylor2025}, but less extreme than what was derived for the more-pristine BH* observed in \citet{naidu25}. That being said, we caution that with photometric data only, these parameters may not be truly constrained and that in particular, the relative strengths of the host galaxy and the BH* are highly degenerate. 

As mentioned in section~\ref{sed}, our \texttt{BEAGLE} galaxy fit to the LRD component also yields an acceptable fit with a DSFG template. With a best-fit reduced $\chi^2=8.7$, this is not as good a fit as the LRD model and in particular struggles to reproduce the rest-frame UV fluxes see Fig.~\ref{fig:SED galaxy fits}. This scenario also requires a much higher stellar mass ($\log(M_{\star}/\mathrm{M}_{\odot})=8.9\pm0.2$) and SFR ($\log(\psi/\mathrm{M}_{\odot}\,\mathrm{yr}^{-1})=1.35_{-0.05}^{+0.03}$), and a more extreme dust attenuation of $A_V=2.0_{-0.1}^{+0.2}$. In addition to the higher $\chi^2$, and while it cannot be completely ruled out at present, we therefore consider the DSFG model less likely than the LRD solution because (i) such a high stellar mass combined with our size measurement in section~\ref{size} yields an extreme density ($>10^5\,\mathrm{M}_{\odot}\,\mathrm{pc}^2$), the same order of magnitude as the densest stellar structures \citep[e.g.,][]{vanzella23,baggen23,Adamo2024}, and (ii) \citet{Knudsen2016} did not detect any dust continuum from this source with ALMA ($<55\,\mu$Jy at $5\sigma$), essentially ruling out SFR greater than 0.5 $\mathrm{M}_{\odot}$/yr. 

The combination of the available observational evidence therefore supports the interpretation of A383-LRD1 as a LRD: Its measured colors (section~\ref{phot}) fall within the range characteristic of previously identified LRDs \citep[e.g.,][]{greene24,kokorev24a,Kocevski2025}, while the size analysis (section~\ref{size}) shows the source to be compact as required -- in fact, consistent with a point-like morphology, and our SED analysis (section~\ref{sed}) favors a model that combines a BH* model with a substantial contribution from a host galaxy. In that respect note that only a small fraction of LRDs were found to show an underlying host \citep[e.g.][]{Chen2025host}. Our analysis, however, remains limited by the absence of rest-frame optical spectroscopy capable of resolving the Balmer lines. A definite conclusion on the LRD (and AGN) nature of A383-LRD1, and the relative significance of its host galaxy, will therefore require spectroscopic observations with JWST to detect its rest-frame optical emission lines, and investigation of their widths and absorption profiles in order to determine the LRD's gas properties. If confirmed, this would make A383-LRD1 one of the very few multiply imaged LRDs discovered to date. With its more evolved host galaxy than the `pristine' LRDs, e.g. the BH* from \citet{naidu25} or A2744-QSO1 \citep[e.g.,][]{furtak24b,Maiolino2025b}, and the high magnifications (see Tab.~\ref{tab:fluxes_all_combined}), A383-LRD1 represents a compelling laboratory to study the interaction of a BH* with its host galaxy and the neighboring blue companion (see section~\ref{sec:BC_discussion}).

\subsection{The blue companion} \label{sec:BC_discussion}
One of the most intriguing aspects of A383-LRD1 is undoubtedly its prominent blue companion, observed $0.3\arcsec$ away in image~A, which corresponds to a projected distance of $\sim380$\,pc in the source plane (section~\ref{lensing}). As demonstrated in section~\ref{size}, this object is marginally resolved in the redder bands and thus also somewhat enlarged, or stretched by the magnification, suggesting that it is intrinsically (slightly)
larger than the LRD at least in the optical -- although it could be smaller than that in the UV (section~\ref{size}). Given its stronger rest-frame UV emission, this object is in fact the one that was previously detected in HST imaging and targeted with DEIMOS and XSHOOTER \citep[see section~\ref{sec:analysis};][]{Richard2011,Stark2015}. It is therefore reasonable to conclude that the UV emission lines detected, Ly$\alpha$ ($\mathrm{EW}_0=138$\,\AA) and C\,\textsc{iii}]$\lambda1909$ \ \AA\ ($\mathrm{EW}_0=22$\,\AA), also originate from the blue component. We nonetheless note that the spatial resolution of neither instrument is sufficient to resolve this distance (in particular if observed by a long slit that is not necessarily aligned with the two objects). For Ly$\alpha$, which was found to be very prominent, we can also invoke the fact that no emission is seen at the position of the LRD in the F814W band, where Ly$\alpha$ should fall at $z=6.027$, to support the line coming from the blue companion (see also intro of section \ref{sec:analysis}). The emission detected from C\,\textsc{iii}] is however not sufficiently strong to invoke the same argument and thus we conclude that at this stage we cannot definitively tell where these UV emission lines come from (see further discussion below).

Our \texttt{BEAGLE} fit to the blue companion's photometry (see section~\ref{sed} and the bottom panel of Fig.~\ref{fig:SED galaxy fits}) achieves an excellent fit (reduced $\chi^2=0.53$), yielding a low stellar mass ($\log(M_{\star}/\mathrm{M}_{\odot})=7.10\pm0.04$), an extremely young age ($\log(t_{\mathrm{age}}/\mathrm{yr})=6.51_{-0.06}^{+0.05}$), and very low dust attenuation ($A_V=0.13\pm0.02$). Surprisingly, the resulting metallicity is rather high ($\log(Z_{\mathrm{gas}}/\mathrm{Z}_{\odot})=-0.40_{-0.08}^{+0.07}$)\footnote{While \texttt{BEAGLE} in principle allows to fit both the stellar and the gas-phase metallicity separately, in this case they are assumed to be the same. Since the stellar metallicity is not constrained by photometry at all, the fitted value is in this case driven by the nebular properties and this represents the gas-phase metallicity.}, which is however unlikely given the high C\,\textsc{iii}] EW. This is also very different from what \citet{Stark2015} found -- an old stellar population with very low metallicity, due to the strong C\,\textsc{iii}] together with the \textit{Spitzer}/IRAC detections. Note the previously reported \textit{Spitzer}/IRAC excess, we can now see originated from both the LRD and the blue companion, contributing equally to the F356W (and thus \textit{Spitzer's} 3.6 $\mu$m) band, but with the LRD governing the flux in the F444W (or \textit{Spitzer's} 4.5 $\mu$m) band. Nonetheless, from the JWST photometry of the blue companion (Tab.~\ref{tab:fluxes_all_combined} and Fig.~\ref{fig:SED galaxy fits}) it is now apparent that its rest-frame optical continuum is very low, and that the F356W and F444W excesses seen in it (please see also Fig. \ref{fig:stamps}) are likely caused by extremely strong H$\beta$+[O\,\textsc{iii}] (EW$_{0}$ of thousands of \AA), and H$\alpha$ (EW$_{0}$ of hundreds of \AA) emission lines, respectively. These strong emission lines seem to drive the high gas-phase metallicity in the fit (though see discussion further below), as well as the extremely young age. 

The fit is entirely dominated by nebular emission, suggesting a young gas cloud perhaps in its very first duty cycle of star-formation. If the gas, however, is just beginning to form stars, then how did it get enriched in metals in the first place? As seen in section~\ref{sec:LRD_discussion}, the LRD's host galaxy seems to be in an evolved state and has undergone several cycles of star-formation. This might suggest a scenario where the gas that makes-up the blue companion was ejected from the LRD's host and is now forming stars on its own. Alternatively, in the case of an uneven coverage of the LRD's AGN by the BH* hydrogen gas shell, the escaping AGN radiation could power the extreme emission lines seen in the blue companion by photo-ionizing the gas, similar to a type~2 AGN, as suggested in e.g.\ \citet{Tang2025}, \citet{Lambrides2025}, or \citet{Chen2025}. In that case, the metallicity would be lower and in agreement with the measured C\,\textsc{iii}]$\lambda\lambda1907,1909$\AA\ EW \citep[e.g.,][]{Stark2015}. Given the frequency of blue companions close to LRDs \citep[e.g.,][]{Chen2025,naidu25}, this might indeed be a common mechanism in the interplay between the BH*/LRD, host galaxy, and their immediate environment. 

As another explanation for the relatively high metallicity obtained in the fit for the blue companion, it is possible that due to its compactness (see section~\ref{size}) the ionization environment in this object is different than what the standard \texttt{BEAGLE} templates can reproduce. In a compact but low-density cloud, even a moderate amount of star-formation can result in strong ionization and thus power the extreme [O\,\textsc{iii}] emission at low metallicity \citep[e.g.][]{Amorin2014,Liu2022}.
As in \citet{Solimano2025}, the weak emission spotted right under the LRD in the F356W (Fig. \ref{fig:stamps}) can suggest some more [O\,\textsc{iii}] emission arising from a gaseous envelope around the LRD, favoring further this scenario. Indeed, the physical separation of $\sim 400$ pc between the LRD and the blue companion may suggest that we are seeing a merger between the two components (or that they both belong to the same underlying galaxy, or host); such multi-component objects have been observed up to $z\approx10$ with similar separations in the source plane \citep[e.g.][]{Hsiao2023,Hsiao2024merger}. It might also be interesting to examine how the SED fit to the blue component, and its resulting metallicity, would look with the inclusion of AGN templates to the fit. We defer, however, such further analysis to future work, possibly when spectroscopic data become available. 

Ultimately we again mention that any conclusions drawn from our photometric analysis remain speculative, and that spatially and spectroscopically resolved data of A383-LRD1 are needed to help reveal the true nature of both the LRD and its blue companion. By directly mapping the emission lines, we would be able to constrain the dynamics of this system (e.g.\ is gas out- or in-flowing to the LRD?) and determine the actual emission line strength and gas metallicity of both components. Currently, and for the foreseeable future, only the \textit{Near Infrared Spectrograph} \citep[NIRSpec;][]{jakobsen22,boeker23} IFU instrument \citep[][]{boeker22} is capable of achieving this at the required wavelengths and both spectral and spatial resolutions.

\subsection{The ALMA [C\,\textsc{ii}] detection} \label{sec:ALMA}

\begin{figure}
    \centering
    \includegraphics[width=0.5\textwidth]{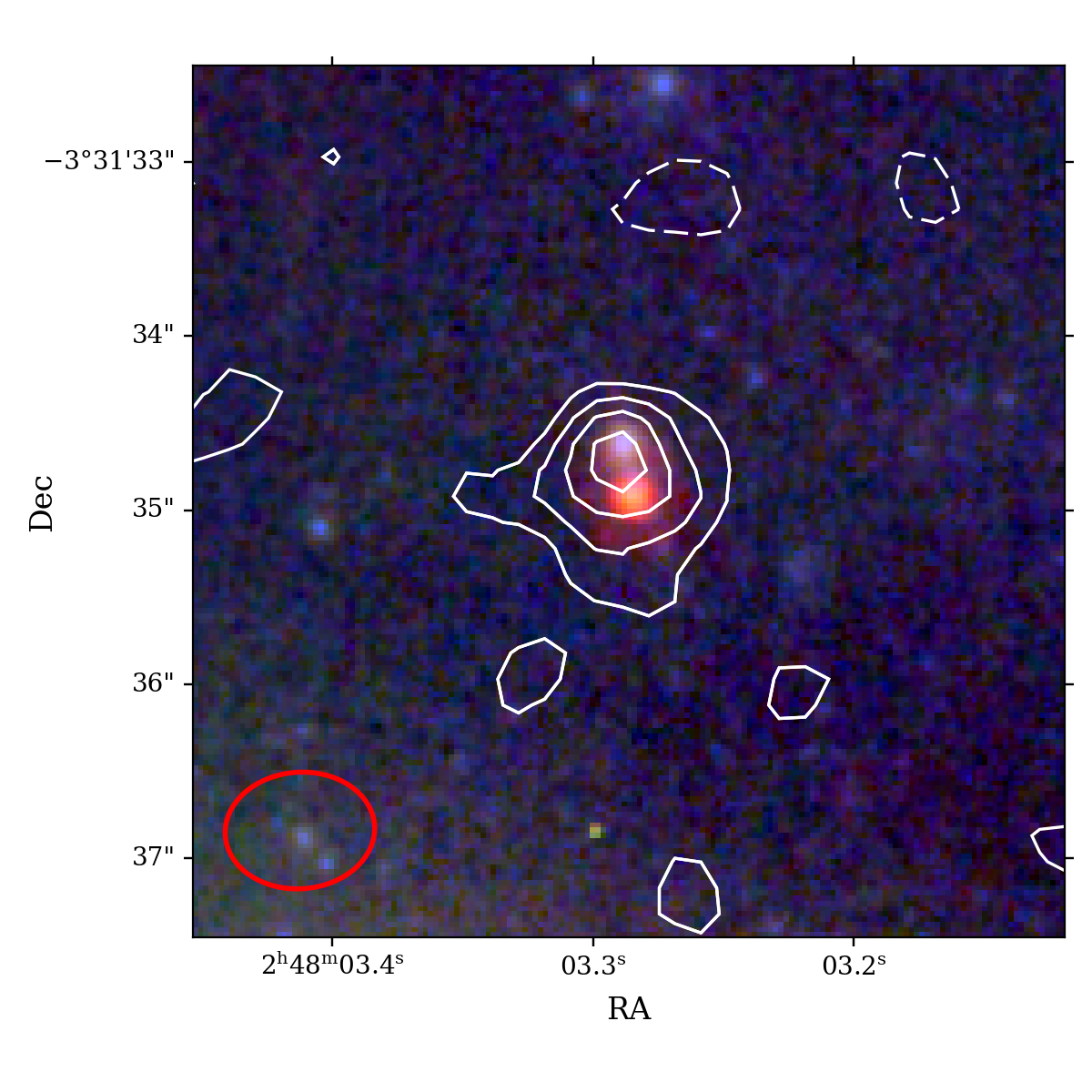}
    \caption{Contours of the ALMA [C\,\textsc{ii}]$\lambda158 \ \mu$m detection from \citet{Knudsen2016}, overlaid over the JWST color image of A383-LRD1. Contours show 2, 3, 4, 5$\sigma$ of the integrated spectral line, and -2$\sigma$ are shown as a dashed line. The obtained resolution is presented as a red ellipse on the bottom left. The contours are centered slightly off the blue companion, with the LRD within the 4$\sigma$ contour.} 
    \label{fig:contoursCII}
\end{figure}

Another piece of the puzzle currently available, is the [C\,\textsc{ii}]$\lambda158\ \mu$m emission, detected thanks to lensing magnification, with an integrated line intensity of $I=0.10\pm0.03\,\mathrm{Jy\,km\,s}^{-1}$ \citep{Knudsen2016}. The [C\,\textsc{ii}] emission is more extended and seen slightly offset from the HST detection, i.e. from the blue companion (see Fig. \ref{fig:contoursCII}). This might suggest that this [C\,\textsc{ii}] emission does not originate from the blue companion alone, but rather also from the extended circum-galactic medium (CGM), or even the LRD itself. Neither of these scenarios is surprising, since [C\,\textsc{ii}]$\lambda158 \ \mu$m is a major cooling line for cold interstellar gas and thus a prime tracer for star-formation, and both the blue companion and the host galaxy of the LRD are actively forming stars. It is even possible that both components are embedded in a common envelope of CGM and actively exchanging gas (and potentially supporting the merger scenario). To date, there has been a dearth of far-infrared emission lines detection from LRDs \citep[e.g.,][]{furtak23d,labbe25,Akins2025,setton25,Xiao2025A&A...700A.231X}. While there has been a [C\,\textsc{i}](2–1) detection in one LRD \citep{Akins2025}, the [C\,\textsc{ii}]$\lambda158\mu$m line from this system, A383-LRD1, to our knowledge, may be the first case of [C\,\textsc{ii}] detected from a LRD to date.

In principle, ALMA is also capable of achieving spatial resolutions rivaling JWST's. It would therefore also be possible to spatially resolve the origin of the [C\,\textsc{ii}] emission, and perhaps detect other far-infrared (FIR) lines in A383-LRD1 with targeted, deep, high-resolution ALMA follow-up observations. Such observations with JWST and ALMA might also allow for a dynamical measure of the black hole mass of the LRD.

\section{Summary} \label{sec:summary}
In this work, we report the discovery of a new, doubly-imaged LRD candidate at $z_{spec}=6.027$ behind the SL cluster Abell~383, which we dub A383-LRD1. The object was previously known as a multiply-imaged high-redshift object from HST imaging, ground-based spectroscopy, and strong \textit{Spitzer}/IRAC and ALMA [C\,\textsc{ii}]$\lambda158\mu$m detections. With the new JWST/NIRCam imaging from the cycle 4 VENUS program, the source is now revealed as a two-component system: a red point source, consistent with typical LRD selection criteria both in terms of colors and size, and a distinct blue companion 0.3\arcsec\ ($380$\,pc) away. Thanks to the relatively high magnification ($\mu=16.2\pm1.2$) derived from a new and dedicated SL model, we place a strong upper limit on the intrinsic size of the LRD of $r<40$\,pc, and measure a very low rest-frame UV luminosity $M_{\mathrm{UV}}=-16.8\pm 0.3$, despite the bright observed magnitudes, e.g. $m_{\mathrm{F444W,A}}=27.2\pm 0.2$. A detailed SED-modeling analysis with \texttt{BEAGLE} and \texttt{bagpipes} reveals the LRD candidate to be comprised of a relatively faint BH* AGN ($\log(L_{5100}/\mathrm{L}_{\odot})=43.2\pm0.2$) and a significant host galaxy ($\log(M_{\star}/\mathrm{M}_{\odot})=7.7_{-0.3}^{+0.5}$). The blue companion on the other hand is consistent with an extremely young ($\log(t_{\mathrm{age}}/\mathrm{yr})=6.51_{-0.06}^{+0.05}$) nebular emission dominated object. We speculate that it could perhaps be a metal-enriched ejecta from the LRD host galaxy, or a young nebula that is photo-ionized by the nearby AGN in the LRD. In the absence of rest-frame optical spectroscopy, this however remains speculative.

With its high magnification and two multiple images, A383-LRD1 provides a unique laboratory to study the interaction of a LRD with its host galaxy, nearby blue companion, and environment. Future follow-up observations with the JWST/NIRSpec IFU could possibly not only detect and confirm its broad Balmer lines, constraining the BH mass, but also spatially resolve and disentangle the strong emission lines from both the LRD and its blue companion. Such observations will be needed to provide more definite answers as to their exact nature and relationship. Moreover, A383-LRD1 is the first LRD system to date with a [C\,\textsc{ii}]$\lambda158\mu$m detection and thus ideally suited for deep high-resolution ALMA follow-up to map the origin of this emission line and search for other FIR lines, and potentially provide a dynamical mass estimate of the BH in the system. Finally, we note that while do not currently detect continuum variability with the modest gravitational time-delay between the two images, $\Delta t_{\mathrm{grav}}=5.20\pm0.14\,\mathrm{yr}$, this inherent time delay could facilitate future variability measurements, and possibly even reverberation mapping campaigns.

\begin{acknowledgements}
      A.Z. acknowledges support by the Israel Science Foundation Grant No. 864/23. L.J.F., H.A., and V.K. acknowledge support from the University of Texas at Austin Cosmic Frontier Center. S.F. acknowledges support from the Dunlap Institute, funded through an endowment established by the David Dunlap family and the University of Toronto. R.A. acknowledges support of Grant PID2023-147386NB-I00 funded by MICIU/AEI/10.13039/501100011033 and by ERDF/EU, and  the Severo Ochoa award to the IAA-CSIC CEX2021-001131-S and from grant PID2022- 136598NB-C32 ``Estallidos8''. F.E.B. acknowledges support from ANID-Chile BASAL CATA FB210003, FONDECYT Regular 1241005, ECOS-ANID ECOS240037, and Millennium Science Initiative, AIM23-0001.
      M.B. acknowledges support from the ERC Grant FIRSTLIGHT, and Slovenian national research agency ARIS through grants N1-0238 and P1-0188.
      E.V. and M.M. acknowledge financial support through grants INAF GO Grant 2022 “The revolution is around the corner: JWST will probe globular cluster precursors and Population III stellar clusters at cosmic dawn”, INAF GO Grant 2024 “Mapping Star Cluster Feedback in a Galaxy 450 Myr after the Big Bang” and by the European Union – NextGenerationEU within PRIN 2022 project n.20229YBSAN - Globular clusters in cosmological simulations and lensed fields: from their birth to the present epoch.
      P.D. warmly acknowledges support from an NSERC discovery grant (RGPIN-2025-06182). Y.H. acknowledges support from the Japan Society for the Promotion of Science (JSPS) Grant-in-Aid for Scientific Research (24H00245) and the JSPS International Leading Research (22K21349). K.K. acknowledges support from the ERC synergy grant 101166930 (RECAP). G.E.M. acknowledges the Villum Fonden research grants 37440 and 13160. M.M. acknowledges financial support through grants INAF GO Grant 2022 ``The revolution is around the corner: JWST will probe globular cluster precursors and Population III stellar clusters at cosmic dawn'' and by the European Union – NextGenerationEU within PRIN 2022 project n.20229YBSAN - Globular clusters in cosmological simulations and lensed fields: from their birth to the present epoch. R.P.N. is grateful for the generous support of Neil and Jane Pappalardo through the MIT Pappalardo Fellowship. M.N. acknowledges support from KAKENHI Grant Nos. 25KJ0828 through Japan Society for the Promotion of Science (JSPS). Y.J-T. acknowledges financial support from the State Agency for Research of the Spanish MCIU through Center of Excellence Severo Ochoa award to the Instituto de Astrof\'isica de Andaluc\'ia CEX2021-001131-S funded by MCIN/AEI/10.13039/501100011033, and from the grant PID2022-136598NB-C32 Estallidos and project ref. AST22-00001-Subp-15 funded by the EU-NextGenerationEU. E.V. acknowledges financial support through grants INAF GO 2024 ``Mapping Star Cluster Feedback in a Galaxy 450 Myr after the Big Bang'' and by the European Union – NextGenerationEU within PRIN 2022 project n.20229YBSAN - Globular clusters in cosmological simulations and lensed fields: from their birth to the present epoch. R.A.W. acknowledges support from NASA JWST Interdisciplinary Scientist grants NAG5-12460, NNX14AN10G and 80NSSC18K0200 from GSFC.
      
      This work is based on observations made with the NASA/ESA/CSA JWST, and the NASA/ESA HST. The data were obtained from the \texttt{Mikulski Archive for Space Telescopes} (\texttt{MAST}) at the \textit{Space Telescope Science Institute} (StScI), which is operated by the Association of Universities for Research in Astronomy (AURA), Inc., under NASA contract NAS~5-03127 for JWST. These observations are associated with the JWST program GO-6882, and the HST program GO-12065. The authors acknowledge the use of the Canadian Advanced Network for Astronomy Research (CANFAR) Science Platform operated by the Canadian Astronomy Data Center (CADC) and the Digital Research Alliance of Canada (DRAC), with support from the National Research Council of Canada (NRC), the Canadian Space Agency (CSA), CANARIE, and the Canadian Foundation for Innovation (CFI). The Cosmic Dawn Center (DAWN) is funded by the Danish National Research Foundation under grant DNRF140.
      
      This research made use of \texttt{Astropy},\footnote{\url{http://www.astropy.org}} a community-developed core Python package for Astronomy \citep{astropy13,astropy18} and \texttt{Photutils}, an \texttt{Astropy} package for detection and photometry of astronomical sources \citep{photutils_v1.13.0}, as well as the packages \texttt{NumPy} \citep{vanderwalt11}, \texttt{SciPy} \citep{virtanen20}, \texttt{Matplotlib} \citep{hunter07}, \texttt{emcee} \citep[][]{foreman-mackey13}, and the \texttt{MAAT} Astronomy and Astrophysics tools for \texttt{MATLAB} \citep[][]{maat14}.
\end{acknowledgements}

\bibliographystyle{aa} 
\bibliography{references} 

\end{document}